\documentclass[twocolumn,superscriptaddress,amsmath,amssymb,pra]{revtex4-2}
\usepackage[utf8]{inputenc}
\usepackage[T1]{fontenc}
\usepackage{graphicx}
\usepackage{dcolumn}
\usepackage{appendix}
\usepackage{bm}
\usepackage{ulem}
\usepackage[usenames,dvipsnames]{xcolor}

\usepackage{subfigure}
\usepackage{bbm}
\usepackage{enumerate}

\usepackage[
  bookmarks=true,
  colorlinks,
  linkcolor=black,
  urlcolor=black,
  citecolor=black,
  plainpages=false,
  pdfpagelabels,
  final,
  breaklinks=true
]{hyperref}

\usepackage{titlesec}

\usepackage{array}

\usepackage{physics}

\DeclareUnicodeCharacter{2212}{-}
\begin{document}
\title{Attosecond spectroscopy using vacuum-ultraviolet pulses emitted from laser-driven semiconductors}

\author{A. Nayak}
 \affiliation{ELI-ALPS, ELI-Hu Non-Profit Ltd., Wolfgang Sandner utca 3., H-6728 Szeged, Hungary}

\author{D. Rajak}
\affiliation{ELI-ALPS, ELI-Hu Non-Profit Ltd., Wolfgang Sandner utca 3., H-6728 Szeged, Hungary}

\author{B. Farkas}
\affiliation{ELI-ALPS, ELI-Hu Non-Profit Ltd., Wolfgang Sandner utca 3., H-6728 Szeged, Hungary}

\author{C. Granados}
\affiliation{Department of Physics, Guangdong Technion-Israel Institute of Technology, 241 Daxue Road, Shantou, Guangdong, China, 515063}
\affiliation{Technion-Israel Institute of Technology, Haifa, 32000, Israel}
\affiliation{Guangdong Provincial Key Laboratory of Materials and Technologies for Energy Conversion, Guangdong Technion-Israel Institute of Technology, 241 Daxue Road, Shantou, Guangdong, China, 515063}

\author{P. Stammer}
\affiliation{ICFO-Institut de Ciencies Fotoniques, The Barcelona Institute of Science and Technology, 08860 Castelldefels (Barcelona), Spain}
\affiliation{Atominstitut, Technische Universit\"{a}t Wien, 1020 Vienna, Austria}

\author{J. Rivera-Dean}
\affiliation{ICFO-Institut de Ciencies Fotoniques, The Barcelona Institute of Science and Technology, 08860 Castelldefels (Barcelona), Spain}

\author{Th. Lamprou}
\affiliation{Foundation for Research and Technology-Hellas, Institute of Electronic Structure \& Laser, GR-70013 Heraklion (Crete), Greece}

\author{K. Varju}
\affiliation{ELI-ALPS, ELI-Hu Non-Profit Ltd., Wolfgang Sandner utca 3., H-6728 Szeged, Hungary}

\author{Y. Mairesse}
\email{yann.mairesse@u-bordeaux.fr}
\affiliation{Universit\'{e} de Bordeaux-CNRS-CEA, CELIA, UMR5107, Talence, France}

\author{M. F. Ciappina}
\email{marcelo.ciappina@gtiit.edu.cn}
\affiliation{Department of Physics, Guangdong Technion-Israel Institute of Technology, 241 Daxue Road, Shantou, Guangdong, China, 515063}
\affiliation{Technion-Israel Institute of Technology, Haifa, 32000, Israel}
\affiliation{Guangdong Provincial Key Laboratory of Materials and Technologies for Energy Conversion, Guangdong Technion-Israel Institute of Technology, 241 Daxue Road, Shantou, Guangdong, China, 515063}

\author{M. Lewenstein}
\email{maciej.lewenstein@icfo.eu}
\affiliation{ICFO-Institut de Ciencies Fotoniques, The Barcelona Institute of Science and Technology, 08860 Castelldefels (Barcelona), Spain}
\affiliation{ICREA, Pg. Llu\'{\i}s Companys 23, 08010 Barcelona, Spain}

\author{P. Tzallas}
\email{ptzallas@iesl.forth.gr}
\affiliation{Foundation for Research and Technology-Hellas, Institute of Electronic Structure \& Laser, GR-70013 Heraklion (Crete), Greece}
\affiliation{ELI-ALPS, ELI-Hu Non-Profit Ltd., Wolfgang Sandner utca 3., H-6728 Szeged, Hungary}

\date{\today}

	\begin{abstract}
		Strongly laser-driven semiconductor crystals offer substantial advantages for the study of many-body physics and ultrafast optoelectronics via the high harmonic generation process. While this phenomenon has been employed to investigate the dynamics of solids in the presence of strong laser fields, its potential to be utilized as an attosecond light source has remained unexploited. Here, we demonstrate that the high harmonics generated through the interaction of mid--infrared pulses with a ZnO crystal leads to the production of attosecond pulses, that can be used to trace the ultrafast ionization dynamics of alkali metals. In a cross--correlation approach, we photoionize Cesium atoms with the vacuum-ultraviolet (VUV) high-harmonics in the presence of a mid-infrared laser field. We observe strong oscillations of the photoelectron yield originating from the instantaneous polarization of the atoms by the laser field. The phase of the oscillations encodes the attosecond synchronization of the ionizing high-harmonics and is used for attosecond pulse metrology. This light source opens a new spectral window for attosecond spectroscopy, paving the way for studies of systems with low ionization potentials including neutral atoms, molecules and solids. Additionally, our results highlight the significance of the source for generating non--classical massively entangled light states in the visible--VUV spectral region. 
	\end{abstract}

\maketitle
 \section*{Introduction}
In the last twelve years, the interaction between semiconductor crystals and intense laser fields has gained significant attention. This interest arises from the combination of advantages offered by many-body physics and ultrafast optoelectronics \cite{Ghimire_NatPhys_2011, Schultze_Nature_2013, Vampa14, Vampa_PRL_2015, Vampa_Nature_2015, Luu_Nature_2015, Wu_PRA_2015, Vampa_PRB_2015, Chang_JOSAB_2016, Kruchinin_RMP_2018, Vampa_NatPhoton_2018, Silva_NatPhoton_2018, Garg_NatPhoton_2018, Ghimire_NatPhys_2019, Juergens_ACSPhoton_2024}. At the core of these investigations lies the process of high harmonic generation (HHG), where the low-frequency photons of a driving laser field are upconverted into photons of higher frequencies. The HHG process is usually induced by the interaction of intense mid-infrared (M--IR) pulses with semiconductor crystals, such as ZnO \cite{Ghimire_NatPhys_2019, Vampa_PRB_2015, Vampa_IEEE_2015}. While this process has been extensively employed to study the electron dynamics in solids in the presence of strong laser fields, its potential for generating attosecond pulses for applications in attosecond science \cite{Agostini_Nobel_2023} has remained experimentally unexploited. Typically, the highest measured harmonics produced by the interaction of semiconductor crystals with intense M--IR (carrier wavelength $\lambda_{L}\sim$ 3.2  $\mu$m) laser pulses lie in the vacuum-ultraviolet (VUV) spectral range with photon energy in the range of few eV \cite{Ghimire_NatPhys_2011}. These photon energies are relatively low compared to the tens of eV photon energies of the harmonics produced by strongly laser-driven atoms \cite{Drake_Hnadbook_2022}. 
The generated pulses are thus below the ionization threshold of the rare gases (which is $I_{P} >12$ eV) used in attosecond pulse characterization techniques (e.g., ref. \cite{Orfanos_APLPhoton_2019} and references therein), preventing their measurement and application. Here, we solve this issue by investigating the photoionization of Cesium (Cs) atoms (with $I_{P(\text{Cs})}\approx3.89$ eV) by a comb of VUV high-order harmonics and a delayed M--IR laser field. We find that the number of electrons ejected along the laser polarization direction oscillates at the frequency of the laser electric field. This modulation arises due to the strong polarization of the Cs atoms by the laser field. The photoelectrons originating from different high-harmonics oscillate almost in phase, indicating synchronization at the attosecond timescale, as predicted by our theoretical calculations conducted using Semiconductor Bloch Equations (SBE). These results establish a novel technique to characterize the attosecond pulses in the VUV range, which employs laser-dressing of the bound electrons rather than the continuum one, as in conventional attosecond metrology \cite{Paul_Science_2001}. 
\\

\begin{figure}	
\centering
		\includegraphics[width=1 \columnwidth]{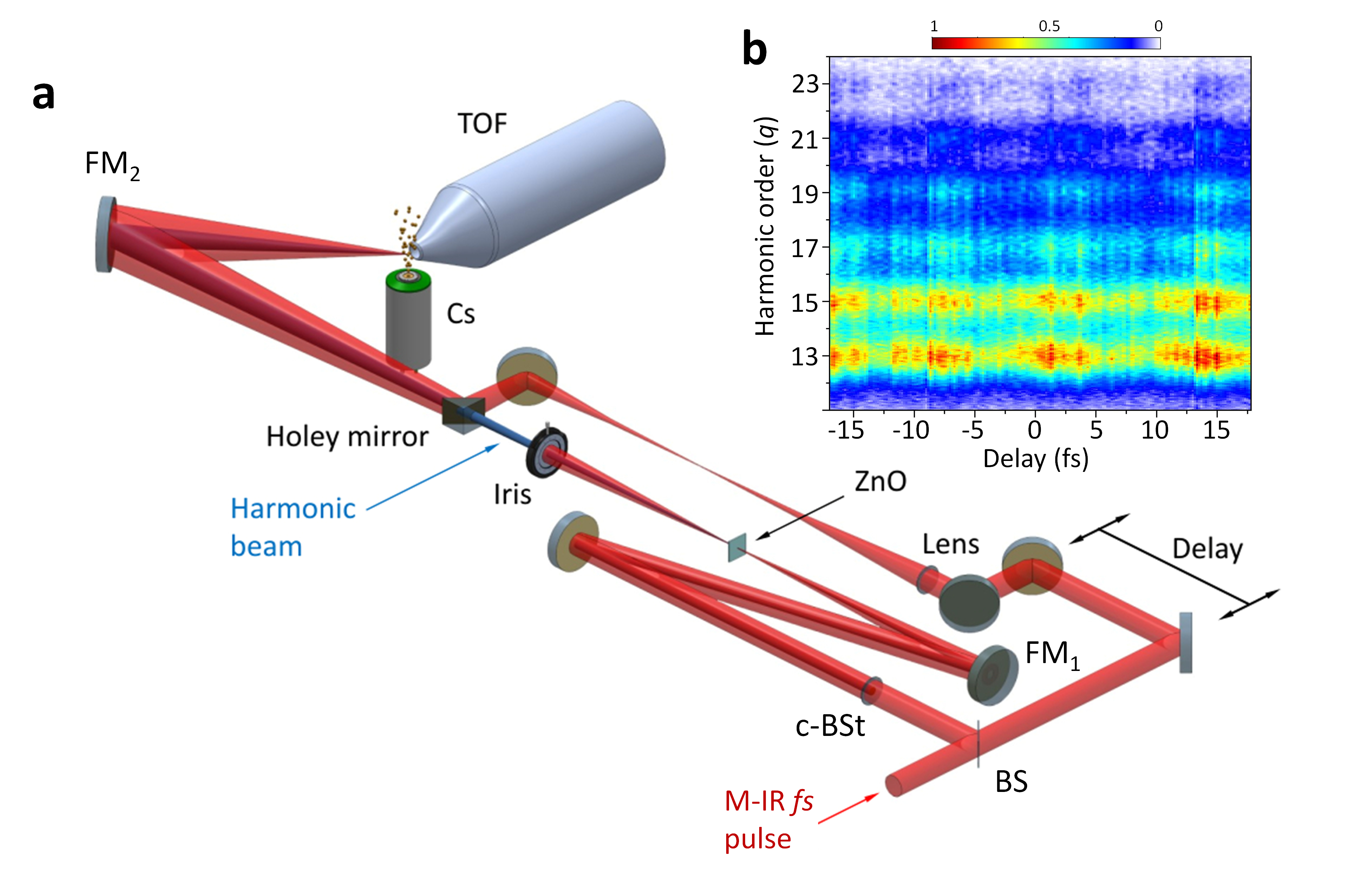}
		\caption{Experimental approach. (a) Cross--correlation method used for the temporal characterization of the harmonics generated by the interaction of intense few-cycle CEP stable M--IR pulses with ZnO crystal. The harmonic and the M--IR fields are spatiotemporally overlapped in the Cs gas. The delay stage was used to control the time delay $\tau$ between the harmonics and the M--IR dressing field. BS and c--BSt, are a M-IR beam separator and a M--IR central beam block, respectively. FM1,2 are focusing mirrors of 50 cm and 15 cm focal length, respectively. The generated photoelectrons are recorded by a $\mu$--metal shielded time--of--flight (TOF) spectrometer. (b) Spectrogram which shows the dependence of the photoelectron spectrum on the delay between the VUV and M-IR fields.}
 \label{fig:setup_scheme}
	\end{figure}

 \section*{High harmonic generation in ZnO}
The experiment has been conducted, at the ELI ALPS facility in the MIR laboratory, using the arrangement shown in Fig.~1a (Supplementary material (SM)). We used a $p$--polarized carrier-envelope-phase (CEP) stable few-cycle M--IR pulses of $\approx 58$ fs duration and carrier wavelength of $\lambda_{L} \approx 3.2$ $\mu$m \cite{Kuhn_JPB_2017}. The M-IR beam entered an interferometer by means of a beam splitter (BS). One arm contains the harmonics generated by the interaction of intense ($\approx 5 \times 10^{11}$ W/cm$^2$) M--IR field with a 500 $\mu$m thick ZnO crystal, and the other, a time delayed ($\tau$) M--IR dressing field. The ZnO optical axis was set perpendicular to the laser polarization in order to generate only the odd harmonic frequencies. The harmonics were recombined with the M--IR at a holey mirror. Both were focused by an Aluminium coated spherical mirror into Cs atoms. The photoelectron spectrum produced by the superposition of VUV and M--IR pulses was recorded by a time-of-flight spectrometer to produce the spectrogram of Fig. 1b. The harmonics contributing to photoionization are those with photons energies $\hbar\omega>I_{P(\text{Cs})}\approx3.89$ eV . Since this energy is above the band gap of ZnO crystal ($E_g\approx 3.2$ eV) the detected harmonics are generated primarily by interband transitions \cite{Vampa_PRB_2015, Vampa_IEEE_2015, Ghimire_NatPhys_2019}. 

To estimate the degree of the synchronization of the emitted harmonics and temporal profile of the VUV pulses produced in the experiment, we performed theoretical calculations using the semiconductor Bloch equations~\cite{Trung16} (SM). The calculated HHG spectrum in the region of interest ($13 \leq q \leq 23$, where $q$ is the harmonic order) (black line in Fig.~2a) shows a reasonable correspondence with the experimental data (Fig.~1b). To investigate if the generated harmonics depict synchronization in the attosecond time scale, we have calculated the harmonic emission times $t_{e}^{(q)}$ (red yellow-filled circles in Fig.~2a) and the temporal structure of the harmonic superposition with $13 \leq q \leq 23$ (Fig. 2b). Due to attosecond phase locking of the emitted harmonics, their superposition results in a VUV temporal profile which consists of a main pulse of $\approx 950$ asec duration, surrounded by secondary pulses reaching 40$\%$ of the main pulse and showing a slightly longer duration. These theoretical results confirm the potential of this source for attosecond spectroscopy.\\

\begin{figure}	
\centering
		\includegraphics[width=0.7 \columnwidth]{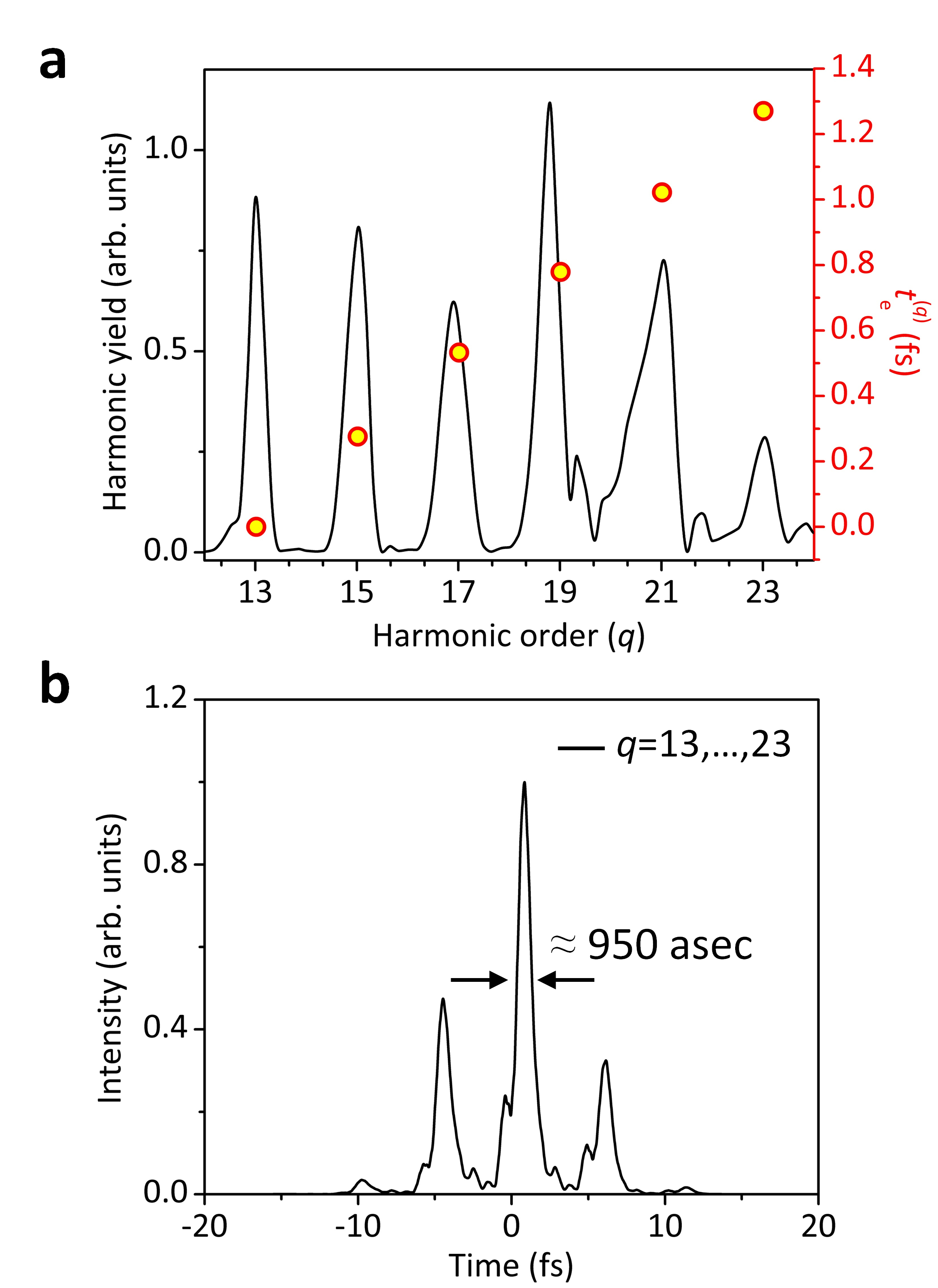}
		\caption{Calculated HHG spectrum, emission times and VUV attosecond pulses. (a) Calculated HHG spectrum (black line) and the corresponding emission times $t_{e}^{(q)}$ (red yellow-filled circles) in the Cs interaction region. In the plot we have set $t_{e}^{(q=13)}=0$. The spectrum is shown in the spectral region used in the pump-probe experiment. (b) Calculated temporal structures of the harmonic superposition with $13 \leq q \leq 23$ resulting to the formation of attosecond pulse trains with the duration of the main pulse in the train to be $\approx 950$ asec.}
 \label{fig:theory:main}
	\end{figure}

 \section*{Attosecond photoionization of Cs}
One of the most important achievements of attosecond light sources relies on their ability to resolve the instantaneous response of atoms \cite{Uiberacker_Nature_2007}, molecules \cite{Neidel_PRL_2013} or solids \cite{Schutze_Science_2014} to the oscillations of the electric field of light. For instance, it has been found that when a small molecule, with polarizability around $20$ atomic units (a.u.), is subjected to a strong IR laser field $(\approx 10^{12}$ W per cm$^2$), the oscillations of the electrons at the laser period induce a modulation of the XUV photoionization cross section, which can be measured using an attosecond pulse train \cite{Neidel_PRL_2013}. Similarly, it has been shown that when argon atoms (polarizability around $10$ a.u.) are polarized by a strong IR field ($\approx 5 \times 10^{12}$ W per cm$^2$) and probed by a single attosecond pulse, the photoelectrons are preferentially ejected in the IR laser field direction \cite{Witting2022}. Here, we show that the VUV pulses synthesized by a superposition of high harmonics generated by semiconductors can be used to track the instantaneous polarization of Cs atoms. The photoelectron spectra shown in Fig. 1b and Fig. 3a consists of peaks corresponding to the absorption of different harmonics. Surprisingly, the spectra does not exhibit sidebands in between the harmonic peaks, that would result from two photon (1 VUV + 1 M--IR ) transitions in the continuum, as is the case in RABBIT measurements\cite{Paul_Science_2001}. The spectra does not show any sign of laser-induced streaking, typical of laser-dressed photoionization with single attosecond pulses \cite{Quere_JMO_2005, Krausz_RevModPhys_2009}. This means that the laser intensity is too weak to induce non-linear transitions in the continuum. However, the overall photoelectron yield presents clear oscillations at the period of the laser field ($T_{0}=10.67$ fs). In other words, more or less electrons are ejected towards the detector, depending on the delay between the VUV and M--IR field. The oscillations of the individual photoelectron peaks are shown in Fig. 3b (raw data as dots, bandpass-filtered data as lines). The electron yield oscillates with a typical modulation depth (SM) in the range of 20$\%$ of the mean (Fig.3b). These results show that the polarizability of Cs (around $400$ a.u. \cite{Schwerdtfeger_MolPhys_2019}) is so high that the pump-probe signal is dominated by the response of the bound electrons to the laser field. This represents a paradigm shift compared to conventional XUV--IR attosecond spectroscopy experiments, where continuum-continuum transitions play a major role. Interestingly, this separation of bound- and continuum- responses to a laser field is a hot topic in surface science, where these two effects are respectively associated with Floquet and Volkov states \cite{Mahmood_NatPhys_2016}.

\begin{figure*}
\centering
		\includegraphics[width=0.8 \textwidth]{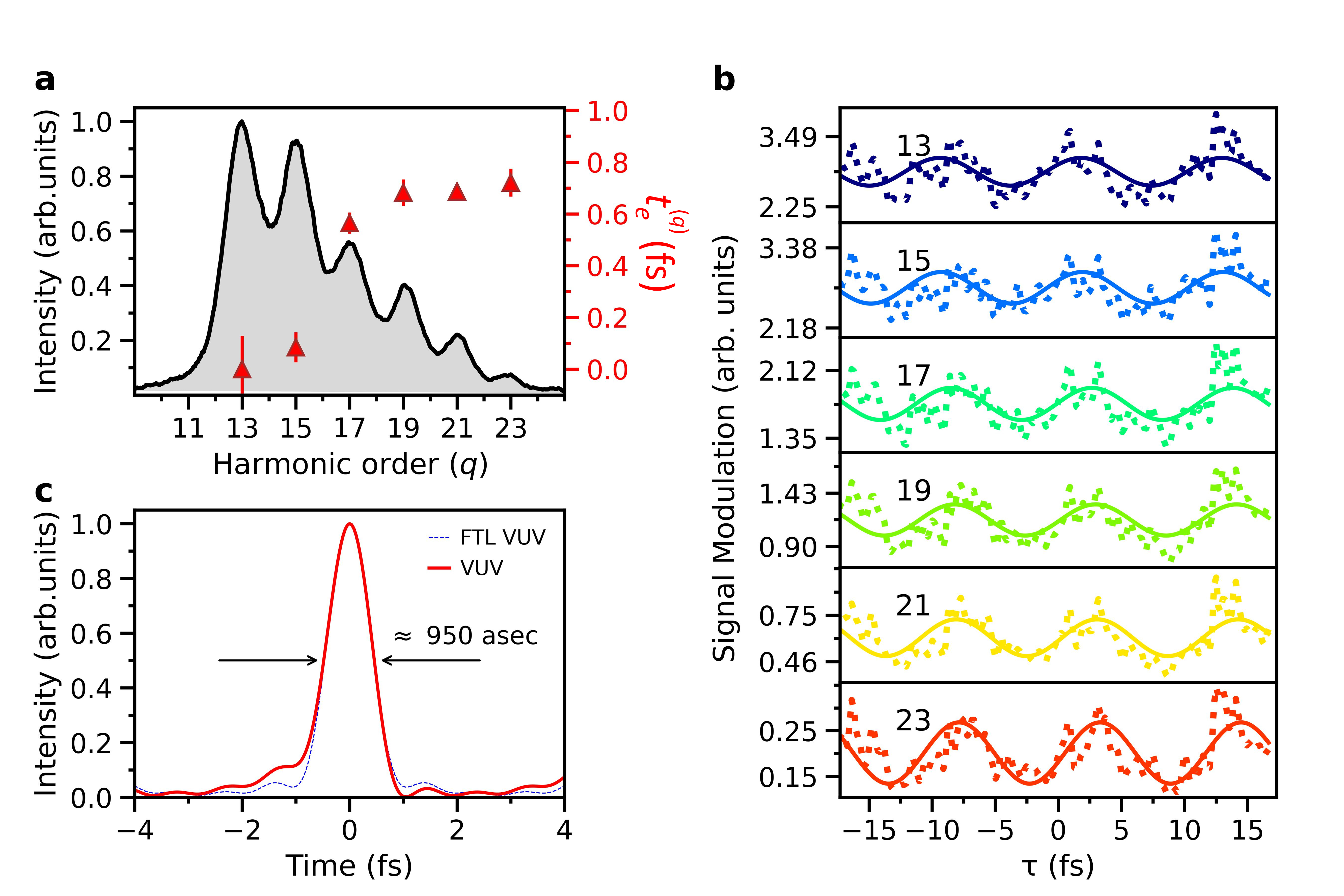}
		\caption{Measured harmonic emission times and VUV attosecond pulses. (a) Measured harmonic spectrum (black line gray--filled area) and the emission times $t_{e}^{(q)}$ (red triangles). In the plot we have set $t_{e}^{(q=13)}=0$. (b) Dependence of the photoelectron harmonic yield on the delay between the VUV and M--IR pulses (raw data as dots, bandpass-filtered data as lines) obtained from the spectrogram in Fig.1b. The photoelectron peaks oscillate with the frequency of the M--IR field $\omega_0$. The shift between the maxima of the harmonic photoelectron yield corresponds to the harmonic emission times (shown in (a)) which have been used to obtain the harmonic phase distribution. (c) The gray depicts the attosecond pulse in the attosecond pulse train in the case of a Fourier Transform Limited (FTL). The red line shows the average attosecond pulse in the attosecond pulse train obtained using the measured phases. The duration of the measured pulse in the train is $\tau_{q=13...23}^{\text{meas.}}\approx 950$ asec.}
 \label{fig:exp:main}
	\end{figure*}

The measurements presented in Figs.~1b and 3a,b also demonstrate that the harmonics produced by ZnO have a temporal structure, able to resolve the instantaneous variation of the atomic polarization. Indeed, we can use the atomic polarization as an amplitude gate to resolve the attosecond synchronization of the different high-order harmonics \cite{Mairesse_Science_2001}. If the harmonics were perfectly synchronized, then their photoelectron signals would maximize at the same delay. Our results show that this is not exactly the case. There is a positive shift of the harmonic photoelectron signal oscillations: the photoelectron yield from higher harmonics maximizes later, which means that the pulses produced in ZnO are positively chirped which is consistent with the results of ref. \cite{Vampa_Nature_2015}. The phase of the photoelectron oscillation provides the emission time $t_{e}^{(q)}$ of the contributing harmonics presented with red triangles in Fig.~3a. The results are in reasonable agreement with the theoretical predictions (red yellow-filled circles in Fig.~2a). The harmonic spectrum and emission times can be used (SM) to retrieve the temporal profile of the average attosecond pulse in the attosecond pulse train. The results show that the attosecond pulses have a $\approx 950$ asec pulse duration, instead of $\approx 910$ asec if there was no chirp (Fig.~3c). This is in good agreement with the theory.

Additionally, we can infer some information on the duration of the attosecond pulse train from the measurements. In the HHG process, an attosecond pulse can be emitted at each laser half-cycle. Thus, in the VUV+M–IR photoionization process, two consecutive attosecond pulses, separated by half a laser period, experience opposite instantaneous polarization by the laser field, cancelling the modulation observed at the laser frequency. This is illustrated in Fig. 4, in which we simulate the effect of an amplitude modulation $\delta (t - \tau)=(1+\delta_{0}E_{\text{M-IR}}(t - \tau))$ induced by the laser field $E_{\text{M-IR}}$ in two-color time-resolved photoelectron spectra. When a single attosecond pulse, with a positive chirp similar to the one measured in our experiment is used (Fig. 4a), the whole electron spectrum oscillates at the laser frequency $\omega_{0}$. The phase of the oscillations reflects the emission time of each frequency component of the spectrum. With a long train of attosecond pulses (Fig. 4b) this oscillation disappears, but side bands, oscillating at twice the laser frequency appear between the harmonic peaks. These side bands correspond to the appearance of Floquet states in the atom. Lastly, we consider the intermediate situation of a short attosecond pulse train, consisting of a dominant central pulse surrounded by weaker satellites as predicted by the theory (Fig. 2). In that case, the photoelectron signal oscillates at frequency $\omega_{0}$ (Fig. 4c), as observed in our measurements. This simple model shows that the observation of the oscillations of the measured signal at the pump laser frequency is an indication that our HHG source produces a main attosecond pulse that can be used for pump-probe attosecond spectroscopy. \\

\begin{figure*}	
\centering
		\includegraphics[width=0.8 \textwidth]{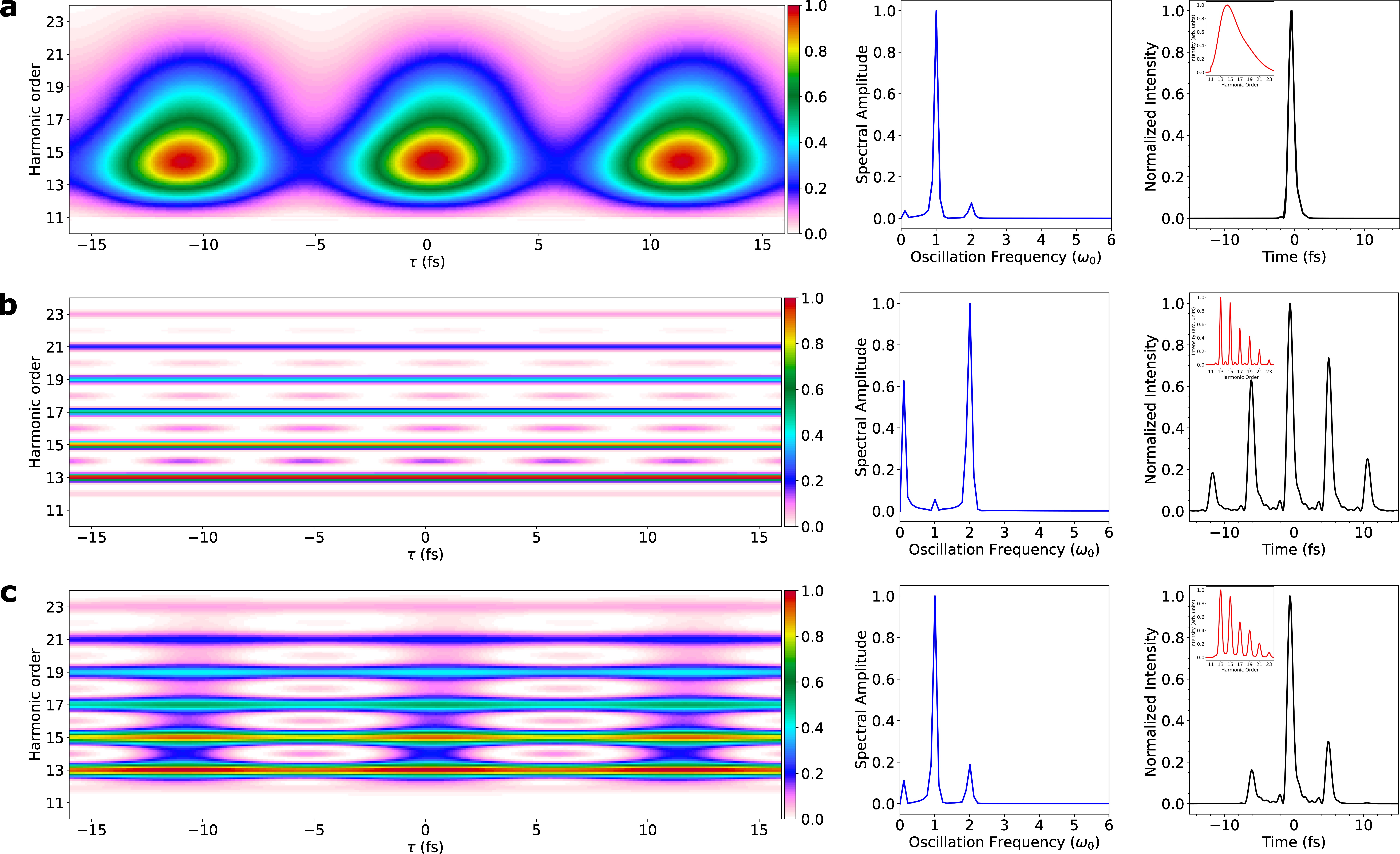}
		\caption{Calculated dependence of the spectrogram on the duration of VUV attosecond pulse train. (a--c) Spectrograms (contour plots), the corresponding frequencies obtained by the Fourier transform along the delay axis of the spectrogram (middle panels), for a single attosecond pulse (a), a long VUV attosecond pulse (b) and an 8 fs long VUV attosecond pulse train (c) (right panels). The insets show the spectrum of the corresponding VUV pulse.}
 \label{fig:exp:main}

\end{figure*}

 \section*{Discussion}
The findings introduce strongly laser-driven semiconductors as a new form of attosecond sources. VUV attosecond pulses are well-suited for investigating the ultrafast dynamics of neutral systems with low ionization energy thresholds. They can be utilized to study the attosecond electron dynamics of excited neutral atoms and the dynamics of coupled electronic and nuclear motion in molecules beyond the Born-Oppenheimer approximation \cite{Nisoli_ROPP_2022}. Additionally, the developed method is suitable for investigating the dynamics of high polarizability systems such as alkali metals \cite{Schwerdtfeger_MolPhys_2019}, as well as solids and surface science \cite{Nisoli_ROPP_2022, Mahmood_NatPhys_2016}. Furthermore, the attosecond synchronization of the coherently emitted multimode harmonic frequency comb underscores the significance of strongly laser-driven semiconductors as a source for generating non-classical, massively entangled light states in the visible to VUV spectral region. Recently, it has been theoretically and experimentally demonstrated that HHG generated by intense laser-atom interactions can be used to produce optical "cat" states, squeezed states, and massively entangled light states \cite{lewenstein_NatPhys_2021, Rivera-dean_PRA_2022, Stammer_PRL_2022, Stammer_PRXQ_2023, Stammer_PRL_2024}. These studies have been conducted in the strong field limit, where high harmonic generation is governed by electron motion in the continuum, with their recombination to the atomic core defining the harmonic emission times. In addition, using fully quantized approaches in strongly laser-driven ZnO crystals, it has been theoretically shown \cite{Javier_PRB_2024, Gonoskov_PRB_2024} that such electron dynamics can also be used to the generation of non-classical and massively entangled states in semiconductors. Hence, the present findings represent a crucial experimental step in this direction, paving the way for numerous applications emerging from the symbiosis of attosecond science and quantum information science \cite{Bhattacharya_ROPP_2023}.

 \section*{Conclusions}
In summary, we have theoretically and experimentally demonstrated that when semiconductor materials are subjected to strong mid-infrared laser fields, they emit attosecond pulses ranging from the visible to the VUV region. We have devised a technique for measuring the temporal synchronization of the generated high-harmonics and the duration of the VUV pulses created by their superposition. The scheme is capable of temporal characterization of the semiconductor HHG, and the produced pulses once characterized provide a new tool for applications in the VUV domain. The method relies on laser-dressed photoionization measurements of Cs atoms, particularly effective in the VUV portion of the harmonic spectrum. Additionally, through this method we have conducted attosecond spectroscopy measurements in Cs atoms. This marks a significant advancement in the study of ultrafast processes and envisage the development of a robust table-top high-repetition rate attosecond light sources using solid state materials.
\\

 \section*{ACKNOWLEDGMENTS}
 
We thank Bálint Kiss, Levente Ábrók and Rajaram Shrestha for their technical support and their efforts on the operation of the mid-IR laser system. We also thank Arnold Péter Farkas, for his excellent work and the methods that he developed for introducing the Cs sample in the interaction chamber. The experiments were carried out at ELI ALPS, and ELI-ALPS is supported by the European Union and co-financed by the European Regional Development Fund  (GINOP-2.3.6-15-2015-00001). 

P. Tzallas group at FORTH acknowledges support from: The Hellenic Foundation for Research and Innovation (HFRI) and the General Secretariat for Research and Technology (GSRT) under grant agreement CO2toO2 Nr.:015922, LASERLABEUROPE V (H2020-EU.1.4.1.2 grant no.871124), The H2020 Project IMPULSE (GA 871161), and ELI--ALPS.

M. Lewenstein group at ICFO acknowledges support from: ERC AdG NOQIA; Ministerio de Ciencia y Innovation Agencia Estatal de Investigaciones (PGC2018--097027--B--I00 / 10.13039 / 501100011033, CEX2019--000910--S / 10.13039 / 501100011033, Plan National FIDEUA PID 2019--106901GB--I00, FPI, QUANTERA MAQS PCI 2019--111828--2, QUANTERA DYNAMITE PCI 2022--132919, Proyectos de I+D+I “Retos Colaboración” QUSPIN RTC 2019--007196--7); MICIIN with funding from European Union Next Generation EU (PRTR--C17.I1) and by Generalitat de Catalunya; Fundació Cellex; Fundació Mir-Puig; Generalitat de Catalunya (European Social Fund FEDER and CERCA program, AGAUR Grant No. 2021 SGR 01452, QuantumCAT \ U16--011424, co-funded by ERDF Operational Program of Catalonia 2014-2020); Barcelona Supercomputing Center MareNostrum (FI--2022--1--0042); EU Horizon 2020 FET--OPEN OPTOlogic (Grant No 899794); EU Horizon Europe Program (Grant Agreement 101080086 — NeQST), National Science Centre, Poland (Symfonia Grant No. 2016/20/W/ST4/00314); ICFO Internal “QuantumGaudi” project; European Union’s Horizon 2020 research and innovation program under the Marie-Skłodowska-Curie grant agreement No 101029393 (STREDCH) and No 847648 (“La Caixa” Junior Leaders fellowships ID100010434 : LCF / BQ / PI19 / 11690013, LCF / BQ / PI20 / 11760031, LCF / BQ / PR20 / 11770012, LCF / BQ / PR21 / 11840013). Views and opinions expressed in this work are, however, those of the author(s) only and do not necessarily reflect those of the European Union, European Climate, Infrastructure and Environment Executive Agency (CINEA), nor any other granting authority. Neither the European Union nor any granting authority can be held responsible for them. 

P. Stammer acknowledges funding from: The European Union’s Horizon 2020 research and innovation programme under the Marie Skłodowska-Curie grant agreement No 847517.

J.~Rivera-Dean acknowledges funding from: the Secretaria d'Universitats i Recerca del Departament d'Empresa i Coneixement de la Generalitat de Catalunya, the European Social Fund (L'FSE inverteix en el teu futur)--FEDER, the Government of Spain (Severo Ochoa CEX2019-000910-S and TRANQI), Fundació Cellex, Fundació Mir-Puig, Generalitat de Catalunya (CERCA program) and the ERC AdG CERQUTE.

Y.~Mairesse acknowledges funding from the Agence Nationale de la Recherche (ANR)--Shotime (ANR-21-CE30-038-01), and thanks Samuel Beaulieu for fruitful discussion. 

M. F. Ciappina and C. Granados acknowledge financial support from the Guangdong Province Science and Technology Major Project (Future functional materials under extreme conditions - 2021B0301030005) and Guangdong Natural Science Foundation (General Program project No. 2023A1515010871).

	\noindent\textbf{{Author contributions:}} A. N., D. R., B. F., C. G.: Equally contributed authors. A. N.: Contributed to the development of the set-up, the experimental runs and the data analysis. D. R.: Contributed to the data analysis. B. F.: Contributed to the preparation of the experiment and experimental runs. C. G.: Contributed to the development of the theoretical approach used for the description of the HHG process. P. S., J. R--D.: Contributed to the theoretical part of the work. Th. L.: Contributed to the experimental part of the work. K. V.: Supported the experiment and contributed to the manuscript preparation. Y. M.: Supervised the spectrogram analysis and the corresponding simulations. M. F. C., and M. L.: Supervised the theoretical part of the work. P. T.: conceived and supervised the project. \\  
	\noindent\textbf{{Correspondence and requests for materials}} should be addressed to Y. Mairesse, M. F. Ciappina, M. Lewenstein, or P. Tzallas.\\

\bibliography{References.bib}{}

  \section*{SUPPLEMENTARY MATERIAL}

{\textbf{Experimental approach:}} The whole optical arrangement and detection units were operated under a pressure of $\sim 10^{-6}$mbar. A beamsplitter (BS) with 70$\%$ reflection and 30$\%$ transmission was used to separate the $\approx 1$ cm diameter incoming M–IR beam in two arms of an interferometer. The reflected beam was passing through a a supper--Gaussian beam block, which creates an annular-shape beam profile, and was focused by a 50 cm focal length mirror into into a 500 $\mu$m thick ZnO crystal. The highest harmonics were produced for a pulse energy of $\approx 25$ $\mu$J corresponding to an intensity $I_{L} \approx 5 \times 10^{11}$ W/cm$^2$. After the ZnO crystal, the harmonic beam, which is spatially confined at the center of the annular--shaped M–IR beam was passing through a 4 mm diameter iris which blocked the annular M-IR beam and let the harmonic beam to pass through. No infrared light could be observed on thermal film past this aperture, indicating the efficiency of the M–IR filtering. In the second arm of the interferometer, the beam was delayed using a translation stage, and recombined with the harmonic beam by a mirror with hole at the center. The divergence of the M--IR beam was matched with that of the harmonics by means of lens. The intensity of the M--IR field was controlled by means of an aperture (not shown) with adjustable diameter placed before the mirror with hole at the center (Fig.1a). The intensity of the M--IR field in the Cs interaction region was roughy estimated to be in the range of $\sim 10^{11}$ W per cm$^2$. Both, the harmonic and M--IR delayed fields, were focused by a 15 cm focal length Aluminum mirror into Cs atoms introduced in the interaction region by means of a specially contracted effusion cell. The generated photoelectrons were recorded by a TOF $\mu$--metal shielded electron spectrometer with the TOF axis to be parallel to the laser polarization. The harmonic spectrum was recorded by measuring the photoelectrons produced by the interaction of the harmonics with Cs atoms in a single--photon ionization process. The signal was found to vanish when a fused silica window was inserted in the beam, confirming the lack of contribution of the M--IR beam to the ionization. Also, no Cs ions were observed when only the M--IR field was focused onto the Cs atoms. This has been measured by operating the TOF in the mass ion spectrometer mode using an ion repeller in the TOF interaction region. The single photon ionization cross section of Cs for $\hbar\omega\gtrsim 5$ eV, varies in the range of $\sim 0.8 \pm 0.2$ Mbarn \cite{Singor_Atoms_2021, Sunitsu_PRA_1983}, such that the photoelectron spectrum can be considered to faithfully reflect the shape of the spectrum of the ionizing radiation. For the pump-probe measurements shown in Fig.3a, the Cs pressure was lowered to avoid polluting the chamber, such that only the lowest, most intense harmonics were detected. Photoelectron spectra were accumulated at each pump--probe delay. The delay step was 0.6 fs. When the Cs pressure was increased at the highest possible level, the photoelectron spectrum exhibits a plateau region extending up to the 31st harmonic (photon energy of $\approx 12$ eV), followed by a cutoff region up to the 43th order ($\approx 16.6$ eV). In this case, and in order to avoid polluting the chamber with Cs, the measurement has been completed within few seconds. For this reason, high Cs pressure conditions have not been used in a pump--probe configuration. The modulation depth discussed in harmonic signal oscillations shown in Fig. 3b, is defined as $M=2(S_{max}-S_{min})/(S_{max}+S_{min})$, where $S_{max}$ and $S_{min}$ are the maximum and minimum values of the oscillating signal, respectively.\\

\textbf{Theory of HHG in ZnO:} Throughout this work, we employ a 1D two-band single-electron semiconductor Bloch equation-based model to tackle the electron dynamics inside the ZnO crystal. This approach has been extensively used for different materials and laser parameters (see e.g., ref. ~\cite{Trung16}). Within this model, it is possible to include, phenomenologically, dephasing and correlation effects through a tunable dephasing time parameter. The valence and conduction bands of ZnO are obtained using a nonlocal pseudopotential approximation (for details, see e.g., ref. ~\cite{Vampa14}). The bandgap energy at the $\Gamma$-point is given by $E_g=0.1213$ a.u. (3.3 eV), in very good agreement with reported experimental values. We use up to 500 $k$-grid points in the $\Gamma$-M direction of the Brillouin zone (BZ) in our simulations, but calculations have also been performed in the $\Gamma$-K direction. The former yields much better agreement with the experimental measurements. The intrinsic chirp of the HHG processes, similar to that observed in atoms, is considered through classical simulations. This enables us to take into account the effects of the attochirp in our theoretical calculations and extract the harmonic phase (see Fig.~2a) (see e.g., ref. ~\cite{Vampa_IEEE_2015}). Our calculations allow for the isolation of the contribution of short electron trajectories by adequately tuning the dephasing time. Setting a dephasing time of 2 fs, we are able to clearly observe the harmonic peaks present in the experimental data. This value is compatible with well-established values~\cite{Vampa14}. Larger values of this parameter make the plateau region of the spectrum noisy and lacking clear harmonic peaks because both short and long electron trajectories contribute, and their coherent sum introduces interference. It can also be confirmed that, for the parameters used in the present work, the intraband contribution above the bandgap is negligible. The harmonic phases have been obtained from the classical theoretical model using the relation $\varphi_{q}=\omega_{q}(t_{e}^{(q)}-t_{E_0})$, where $t_{E_0}$ is the time where the M--IR field $E_0$ takes its maximum value and $t_{e}^{(q)}$ (shown in Fig.2a) is the calculated harmonic emission time.\\

\textbf{VUV pulse reconstruction:} To reconstruct the temporal shape of the VUV pulses, we have used the measured harmonic spectral amplitudes and phase distributions (shown in Fig.3a). The harmonic amplitude distribution has been obtained by integrating along the y-axis the recorded spectrogram shown in Fig. 1b. The oscillations of each harmonic at the laser frequency were analyzed by Fourier-transform. The phase of the oscillation of a given harmonic $q$ provides the emission time $t_e^{(q)}$ of this harmonic within the laser cycle, up to an additive constant that is undertermined in our experiment. This means that we can reconstruct the emitted attosecond pulses, but not their synchronization with the  driving laser field \cite{Dinu_PRL_2003}. The phase of the different harmonics $\varphi_q$ were obtained by integrating the emission times: $t_e^{(q)}=\frac{\varphi_{q+2}-\varphi_q}{\omega_{q+2}-\omega_{q}}$. The intensity profile of the average attosecond pulse in the train was finally obtained by using the harmonic amplitudes $A_q$ and phases $\varphi_q$: $I(t)=\abs{\sum{A_q e^{i\varphi_q}}}^2$.
\\

\textbf{Dependence of the spectrogram of the VUV pulse duration:} To investigate the effect of VUV pulse and pulse train on the photoelectron spectrogram we use the first order perturbation theory under the single active electron approximation to calculate the transition amplitude $\Gamma _{\mathbf{v}}(\tau)$ . The transition amplitude $\Gamma _{\mathbf{v}}(\tau)$ to the final continuum state $\ket{\mathbf{v}}$ with the momenta $\mathbf{v}$ is given by \cite{Quere_JMO_2005},

\begin{equation}
	\Gamma _{\mathbf{v}} (\tau ) = -i \int_{-\infty}^{\infty} dt \mathbf{d}_{\mathbf{v}}(t - \tau) E_{VUV} (t) e^{i S(t, \tau, \mathbf{v})}
\end{equation}
with
\begin{equation}
	S(t, \tau, \mathbf{v}) = \int_{t_i}^{\infty} dt \Bigg( I_p + \frac{(\mathbf{v} + E_{\text{M-IR}}(t - \tau))^2}{2}\Bigg)
\end{equation} 

The recorded photoelectron spectrum does not show any side-bands which is due to low M-IR intensities. We thus aproximate the action $S$ to $S = I_p t +\frac{v^2}{2m}t$. Although the M-IR field strength is weak, it still induces an amplitude modulation in the photoelectron spectra due to the high polarizability of Cs. We model this by introducing an amplitude modulation of the dipole transition matrix element proportional to the M-IR field: $\mathbf{d}_{\mathbf{v}} (t - \tau) = \mathbf{d}_{\mathbf{v}}^0 (1+\delta_{0}E_{\text{M-IR}}(t - \tau ))$, where $\mathbf{d}_{\mathbf{v}}^0$ is the laser field free dipole trnasition matrix element and $\delta_{0}$ is the depth of amplitude modulation.

\end{document}